\documentclass{aa}
\usepackage[dvips]{graphicx}
\usepackage{psfig}

\begin{document}
\title{
First Spectroscopic Evidence for Carbon Stars Outside the Local Group:
Properties of a Massive Star Cluster in 
NGC 7252\thanks{Based on observations obtained at the NTT 3.5m of
the European Southern Observatory, Chile}.}

\author{M.\,Mouhcine\inst{1,}\thanks{present address: Departement of Physics 
           \& Astronomy, UCLA, Math-Sciences Building 8979, Los Angeles, 
	   CA 90095-1562}, A.\,Lan\c{c}on\inst{1}, C.\,Leitherer\inst{2}, 
        D.\,Silva\inst{3}, M.A.T.\,Groenewegen\inst{3,4}}

\institute{
Observatoire Astronomique de Strasbourg, 11 rue de l'Universit\'e, 
F-67000 Strasbourg, France
\and
Space Telescope Science Institute, 3700 San Martin Drive, Baltimore, 
MD 21218, USA
\and
European Southern Observatory, Karl-Schwarzschild-Str.2, D-85748 Garching, 
Germany
\and
Instituut voor Sterrenkunde, PACS-ICC, Heverlee, Belgium  
}

\date{Received ; accepted ..}

\authorrunning{Mouhcine et al.}

\titlerunning{Near-IR Spectrum of Cluster in NGC 7252}

\abstract{
We present near-infrared [$1\,-2.3\,\mu$m] spectroscopy of the massive 
intermediate age star cluster W3 in the merger remnant and proto-elliptical 
galaxy NGC 7252, obtained with the NTT telescope. 
This cluster has an age when the integrated near-infrared properties of 
a stellar population are dominated by the cool and luminous Asymptotic Giant 
Branch (AGB).\\
We compare the data with instantaneous burst model predictions from new 
evolutionary synthesis models that include: (i) the computation of the evolution 
through the thermally pulsing AGB (TP-AGB) for low- and intermediate-massive stars, 
with the initial mass and metallicity dependent formation of carbon stars; 
(ii) spectroscopic data from a new stellar library in which differences between 
static red giants, variable oxygen rich TP-AGB stars and carbon stars are accounted 
for. The new evolutionary model predicts that the contribution of carbon rich stars 
to the luminosities in the near-IR passbands is a strong function of metallicity.\\
The comparison of the data to the models clearly shows that carbon stars are 
present: for the first time, carbon rich star spectral features are thus detected 
directly outside the Local Group galaxies. Good fits to the available 
optical/near-IR photometry and the near-IR spectrum of NGC 7252-W3 are found for 
an age of 300-400\,Myr and A$_{V}\,\simeq\,0.6-0.8$. 
The models show that these parameters depend weakly on the model metallicity in the 
range of Z/Z$_{\odot}=0.4-1$, with higher likelihood for solar metallicity models.\\
At solar metallicity, a mixture of carbon rich and oxygen rich stars is predicted.
The strength of the near-IR molecular bands that originated from oxygen rich 
AGB stars can be used to constrain the absolute T$_{eff}$ scale of these objects,
i.e. a relation between colour and T$_{eff}$. We found that, in the framework
of our set of evolutionary tracks, the data are more 
consistent with the temperature scale calibrated on Long Period Variables than on 
giant stars. At a given colour, variable AGB stars have a lower T$_{eff}$ than 
static (or quasi-static) M giants.
\keywords{stars: AGB -- galaxies: star clusters --
galaxies: stellar content -- infrared: galaxies}
}

\maketitle

\section{Introduction}
For the past decades, great efforts have been devoted to  
the determination of the star formation history of complex
stellar systems, such as galaxies, with the ultimate aim of 
understanding the history of star formation on cosmological
timescales. In such composite systems, we look at a 
mixture of stellar populations with large spreads in 
age and metallicity (M\"oller et al. 1997).
The picture is further blurred by the presence
of dust. An extended observational wavelength coverage is a
major advantage in attempts to break the degeneracies between the 
effects of age, metallicity and extinction on galaxy light.
In this context, near-infrared (near-IR) observations are justified 
by the relatively low sensitivity of near-IR light to dust extinction
(stellar subpopulations that are heavily obscured contribute more to the 
near-IR light than they do at shorter wavelengths),
and because the near-IR light originates specifically in subpopulations
of old or metal-rich stars (Frogel et al. 1978, Rieke \& Lebofsky 1979, 
Frogel 1985, Silva 1996,
Lan\c{c}on et al. 1996, Goldader et al. 1997, Fritze-von Alvensleben
1999). 
In practice, the interpretation of the near-IR stellar energy distribution
frequently remains limited  to the global confirmation of the presence 
of ``old" stars or, in starburst environments, to the suggestion of a 
$\sim 10^7$\,yr young population of red supergiants (Oliva et al. 1995,
Lan\c{c}on \& Rocca-Volmerange 1996). 
In particular, the specification of the actual age of the ``old"
stars seems to have stayed out of reach. This is mainly due to
large uncertainties in the physics of cool stars, used
as input to the population synthesis calculations (Charlot et al. 1996, 
Girardi 1996, Origlia et al. 1999, Lan\c{c}on et al. 1999). 

Important progress in extragalactic near-IR astronomy
would be achieved if contributions from {\em intermediate age} 
($<$2\,Gyr, but $>$\,100\,Myr) and {\em old} ($>$2\,Gyr) populations 
could be safely separated in the integrated light of galaxies. 
At intermediate ages, asymptotic giant branch stars (AGB stars)
are responsible for most of the K-band flux. Integrated broad band 
colours are determined by these stars (Frogel et al. 1990, 
Bruzual \& Charlot 1993, Girardi \& Bertelli 1998, 
Mouhcine \& Lan\c{c}on 2002a). 
But broad band colours remain ambiguous,
as they are similarly affected by extinction or by an additional
population of red supergiants, AGB stars or metal-rich giants.
Lan\c{c}on et al. (1999; see also the early review of 
Lan\c{c}on 1999) have suggested the use of the broad
spectral signatures of upper AGB stars 
for an age separation. The basic idea is that most upper AGB stars 
are Long Period Variables (LPVs), and that this variability produces 
more extended atmospheres, thus leading to significantly stronger 
molecular bands. When oxygen-rich AGB stars become carbon-rich, they 
display specific molecular signatures that can again be identified. 
For the first
time, stellar spectral libraries that included oxygen-rich LPVs
and carbon stars had been used as input to an evolutionary
spectral synthesis code. However, the qualitative predictions are 
strongly dependent on as yet only partly constrained AGB evolution 
parameters (Lan\c{c}on et al. 1999, Mouhcine \& Lan\c{c}on 2002).
For a given set of evolutionary tracks and a given library of 
stellar spectra, a very influential parameter is the temperature
scale used to relate the two. 

To improve the reliability of the synthesis models, we need to 
check the accuracy of their physical ingredients in simple cases.
Star clusters provide an invaluable opportunity to do so.
Because of their proximity and of the large range of 
properties they display, Magellanic Cloud (LMC, SMC) clusters 
were used extensively. However, LMC/SMC 
clusters contain only a handful of luminous red stars. 
Their integrated properties are strongly affected by the 
stochastic character of this small subpopulation (Santos \&
Frogel 1997, Ferraro et al. 1995). Lan\c{c}on \& Mouhcine (2000)
have shown that coeval populations with more than $10^5$\,M$_{\odot}$
of stars are needed if one wishes random deviations from the 
mean near-IR properties predicted by population synthesis models 
to remain smaller than the effects of the model parameters 
of astrophysical interest. Repeated observations of $10^5$\,M$_{\odot}$
clusters or even larger masses are required in order to cancel
effects of the variability of the brightest stars as well. Such 
massive clusters are rare in nearby resolved galaxies (Grebel 2000).

Over the last few years, the {\it{Hubble Space Telescope}} has revealed 
the presence of numerous young, compact and very bright knots in merging 
galaxies, e.g. NGC 4038/4039 (Whitmore \& Schweizer 1995, Whitmore et al 1999), 
NGC 3597 (Holtzman et al. 1996), NGC 7252 (Miller et al 1997) and NGC 1275 
(Carlson et al 1998). Their colours, luminosities, sizes and spatial extent 
agree with them being proto-globular clusters. 
These results suggest that galaxy interaction and mergers are favorable 
environments for the formation of globular clusters. The ``Atoms for 
Peace galaxy" NGC 7252 (Arp 226) is one of the most extensively studied 
merging remnants. 
Star clusters of exceptional luminosity were discovered in NGC\,7252
(Schweizer 1982, Whitmore et al. 1993). The analysis of UBV and 
I photometry leads to estimated ages of several 100 Myr
and extraordinary masses of $\sim 10^7$\,M$_{\odot}$
for the most luminous ones (Miller et al 1997). The NGC 7252 massive 
clusters offer us the possibility to calibrate the stellar input of 
our near-IR modelling of intermediate age SSPs without worrying about 
stochastic fluctuations.

In Section \ref{obs} we describe the observations and the data 
reductions. Section \ref{models.sec} presents an overview of our new 
modelling of near-IR spectra of SSPs. 
In Section \ref{Cstar.evidence.sec}, we highlight the evidence for 
a significant population of carbon stars in he near-IR spectrum of 
NGC\,7252--W3 (hereafter W3). In Section \ref{Age_Ext.sec},
we combine UBVIK$_s$ photometry and the near-IR spectrum to derive 
new constraints on the age of the cluster and the extinction
on the line of sight. Constraints on the cluster metallicity and 
on the temperature scale of AGB spectra are discussed in 
Section \ref{Z_Tscale.sec}. A comparison with previous models, 
in which the evolutionary tracks for the TP-AGB did not account for 
``hot bottom burning", is given in Section \ref{comp.sec}. Finally, 
Section \ref{concl} summarizes the main conclusions. 

\section{Cluster selection, observations and data reductions}
\label{obs}
\subsection{Selection}
\label{selection.sec}

The "Atoms for Peace" galaxy, NGC 7252, is located at 
$\alpha=22^{h}20^{m}44^{s}.8$, $\delta=-24^{\circ}40^{'}42^{"}$ (J2000) 
(Miller et al 1997) with a 
recession velocity relative to the local group of $4828\,{\rm km\,s}^{-1}$ 
(Schweizer 1982). We adopt a distance modulus $(m-M)_{o}=34.08$ and a Milky 
Way foreground extinction of $A_{V}=0.04$ (de Vaucouleurs et al 1990; 
H$_o$\,=\,75\,km\,s$^{-1}$\,Mpc$^{-1}$). 

The candidate clusters were selected from the lists 
of Whitmore et al. (1993) and Miller et al. (1997). Only clusters 
with estimated ages between 0.1 and 1\,Gyr were considered good
candidates for our purpose. At these ages, no red
supergiant stars are present any more, which eliminates 
one potential source of contamination.
Metallicities between Z=0.3\,Z$_{\odot}$ and Z=Z$_{\odot}$ (based on 
the chemical evolution models of Fritze-von Alvensleben \& Gerhard 1994) 
in fact suggest
a much narrower age range of 600 to 800\,Myr for these clusters. 
Schweitzer \& Seitzer (1998) re-analysed the photometry and combined
it with optical spectroscopy. They favoured solar abundances
and slightly younger ages  (300--600\,Myr for cluster W3).
Finally, Maraston et al. (2001) added K band photometry to the data
and used updated models to confirm a metallicity range
of 0.5 to 1\,Z$_{\odot}$, and to restrict the range of ages to
300--500\,Myr (for cluster W3, they favoured Z=0.5\,Z$_{\odot}$
and an age of 250--300\,Myr). 

The observable intermediate age clusters are intrinsically
bright, with M$_{V}\la -14$ and M$_{K}\la -16.5$.
Assuming standard stellar initial mass functions (Salpeter 1955,
or Scalo 1986), this corresponds to stellar masses of the
order of 10$^7$\,M$_{\odot}$ or more: all these clusters 
are massive enough to host a representative population
of luminous AGB stars.

\subsection{Imaging}

J and K$_{\rm s}$ (K short; $\lambda/\Delta\lambda=2.162/0.275\,\mu$m) 
images of NGC 7252 were taken with the IR imager-spectrometer SOFI on the 
European Southern Observatory New Technology Telescope (La Silla, Chile)
between August 14 and August 19, 1999. Two of the five half nights allocated 
to the program were adequate, and were used to acquire the imaging and the 
spectroscopic data of this paper. 

SOFI has a field-of-view of $\sim 5^{'}\,\times\,5^{'}$ and 
a pixel size of $0.292^{''}$. All images were acquired by 
integrating repeatedly for equal amounts of
time on the galaxy and on nearby sky fields. Both galaxy and 
sky frames were dithered with offsets $\ge\,10\,\arcsec$, so that 
a median of the frames would effectively remove field stars, the 
detector defects and cosmic rays.  This technique also has the advantage 
of minimizing residual flat-fielding errors.  The individual object 
integration times were 200\,s in K$_{\rm s}$ ($20\times 10$\,s) and 
80\,s in J ($20\,\times 4$\,s). The object/background integration 
cycle was repeated three times for the K$_{\rm s}$ band 
and 8 times for the J band. The resulting total on-object exposure times 
are 1000\,s in K$_{\rm s}$ and 560\,s in J. Standard stars from Persson 
et al. (1998) were used for the photometric calibration.

\begin{figure}
\centerline{\psfig{figure=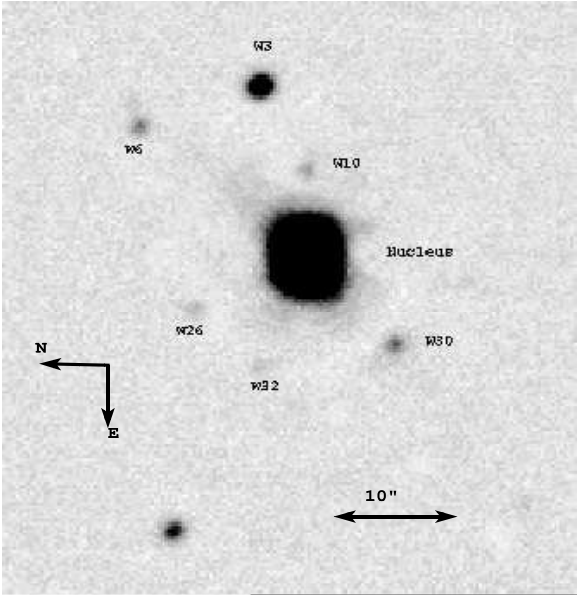,width=8.cm,height=9cm}}
\caption{Masked version of NGC 7252 J band image. Clusters are 
marked following Miller et al 1997.  
 }
\label{Jimage7252}
\end{figure}

The images were processed with IRAF\footnote{The Image Reduction and
Analysis Facility by the National Optical Astronomy Observatories which 
are operated by AURA, Inc., under cooperative agreement with the National 
Science Foundation, USA.}.
A dark frame was first subtracted from every object and sky image of a 
given data set. The dark frame was constructed from a median of dark 
exposures taken at the beginning or at the end of the night. The dark 
currents were very stable. Flatfields in the two bands were made by median 
combining the standard star images and subtracting the corresponding dark 
frame.
These sky flatfields were found to be superior to the dome flatfields.
The flatfield image was divided into each dark subtracted object and sky 
image. A reference background image for each band was then constructed by 
median-combining the sky images. This image was subtracted from all galaxy 
images of a given series of observations. 
The individual sky-subtracted galaxy images were then carefully shifted 
and median combined using a sigma-clipping algorithm to form a single final 
image of NGC 7252.
We modelled the galaxy with a $31\,\times\,31$ pixels median filter and 
subtracted the light profile of the galaxy, in order to obtain a flat, 
low-noise background. The masked J-band image of the central region of the 
galaxy is shown in Fig.\,\ref{Jimage7252}, where the star clusters are 
marked using the identification numbers of Whitmore et al. (1993).

\subsection{Spectroscopy}

The low resolution grisms of SOFI make it possible to cover the
spectral range between 0.95 and 2.52 $\mu$m in only two settings. 
The {\it{blue}} setting ranges from 0.95 and 1.64 $\mu$m , the 
{\it{red}} setting from 1.53 to 2.52 $\mu$m. The extended wavelength 
coverage of each setting and the overlap between the two are decisive 
advantages of SOFI over other available medium resolution infrared 
spectrometers on 8-10m class telescopes. Indeed, the features
that distinguish stellar populations of various ages in the near-IR
are extremely broad (Lan\c{c}on et al. 1999, Mouhcine \& Lan\c{c}on 2000). 
They reach over the telluric absorption bands that
mark the edges of the J, H and K atmospheric windows. The vast majority
of near-IR spectrometers only allow the observation of one window at
a time. Varying slit losses from one observation to the next then make it 
extremely difficult to recover the relative flux levels of the spectral
segments and to measure and use the spectral features.

For the first spectroscopic observations ({\it{blue}} setting), the $1\arcsec$ 
slit was placed across two clusters, W3 and W30 (position angle 117.4). 
This slit position also covers the galaxy emission close to the nucleus. 
As weather made it clear that not enough signal would be collected 
for W30, the subsequent observations were made at position angle 104.6,
with the slit running through W3 and the galaxy nucleus.
The SOFI slit is $290\arcsec$ long. Nodding along the slit 
for first order background subtraction was thus  possible.
Our observations consist of series of 40 minute nodding sequences, amounting 
to 9600\,s exposure in the blue setting and 10800\,s exposure in the red 
settings. Standard star spectra were taken before and after each sequence
on NGC\,7252.

The spectroscopic data for cluster W3
were reduced in the following steps. Pairs of frames obtained at
two positions along the slit were combined by subtraction and flat-fielded,
using the adequate lamp-on -- lamp-off frame. The resulting frames carry
a positive and a negative version of the spectrum. The numerous cosmic rays
could only partially be removed with automatic procedures; careful eye 
inspection was used to remove cosmic rays very close to the cluster and
galaxy light. Then the spectral images were
calibrated in wavelength using a xenon lamp spectrum as a reference.
To extract a one dimensional spectrum for the cluster, the cluster emission 
was traced along the two-dimensional images. The adopted extraction aperture 
width varied between 5 and 6 pixels,
depending on the seeing. The combined galaxy background and residual sky 
background at each wavelength was estimated via a cubic spline fit to 
background windows on either side of the trace center.
Following the background subtraction, each individual spectrum was flux 
calibrated using an average of standard star spectra taken before and 
after the scientific exposure.
The division by the standard star spectrum removed telluric absorption 
features. Kurucz spectra (1993) were used as models for the intrinsic 
energy distribution of the standards, and thus to recover the energy 
distribution of the cluster.  Individually calibrated cluster spectra 
from different images were then combined into one final spectrum. The 
square of the signal-to-noise ratio was used to weight the individual 
spectra. Finally, the {\it{blue}} and {\it{red}} spectra were merged.

\section{Models for the near-IR emission of stellar populations}
\label{models.sec}

Studies of the evolution of the near-IR spectral energy distribution 
of intermediate age stellar populations need to take into account 
stars of the Early and of the Thermally Pulsing AGB, since
both contribute significantly to the near-IR light
(Frogel et al. 1990, Charlot \& Bruzual 1991, Lan\c{c}on 1999).
The sets of evolutionary tracks or isochrones made available 
by stellar theory groups for use in population synthesis models do 
not in general extend beyond the Early AGB. An extension through 
the TP-AGB is required. Unfortunately 
the mechanisms driving the TP-AGB evolution are complex
and still poorly understood (e.g. Iben \& Renzini 1983, Habing 1996, 
Olofsson 1999).  Uncertainties in the physics of mass loss, of
the third dredge-up process, and of hydrogen burning at the bottom of 
the convective envelope, together with the lack of reliable 
models for the molecular spectral features of the coolest stars,
make the inclusion of TP-AGB stars a difficult task. 
The integrated properties of the stellar populations in the near-IR 
will depend on the modelling of the TP-AGB phase (see Girardi \& 
Bertelli 1998 for a clear illustration). 

The analysis of the clusters of NGC\,7252 is based
on the new spectral synthesis models presented by
Mouhcine \& Lan\c{c}on (2002a, hereafter ML2002), 
in which the TP-AGB contribution to the integrated 
light is taken into account. The synthetic stellar evolution code 
for the TP-AGB incorporates the above-mentioned physical 
processes, through parameters that have been adjusted to reproduce
observational constraints from the Magellanic Clouds and the solar
neighbourhood. It predicts the evolution of stars of various masses 
in the HR diagram, as well as the evolution of the surface chemistry
and of the stellar mass (thus the formation of carbon stars and 
of dust-obscured mid-infrared sources). Moreover, it uses a 
stellar library that includes an appropriate optical + 
near-IR spectrum for each of these evolutionary stages.
The reader is referred to ML2002 for a detailed description.
Let us however summarize the main properties here.

Up to the E-AGB, the stars are assumed to follow the 
evolutionary tracks of the Padova group (e.g. Bressan et al. 1993 for
[Z=0.008, Y=0.25], Fagotto et al. 1994 for [Z=0.02, Y=0.28]).
The extensions through the TP-AGB include known physical processes 
that affect the evolution and play a dominant role in determining 
TP-AGB lifetimes, the extent of nuclear processing and the chemical 
surface abundances: the third dredge-up, the envelope burning,
the mass loss and its final superwind phase. 
From the end of the E-AGB on, the total mass, core mass, 
effective temperature, bolometric luminosity and carbon to oxygen ratio
in the envelope are evolved according to semi-analytical prescriptions
(see Wagenhuber \& Groenewegen 1998). The basic relations of the model 
are the following:
\begin{enumerate}
\item a detailed core mass-luminosity relation;
\item the core mass-interpulse relation;
\item the rate of evolution of the core mass;
\item an algorithm to evaluate the effective temperature;
\item a test of the occurrence of CNO burning at the base of the convective 
      envelope;
\item a prescription to evaluate the mass loss rate as function of
      the stellar parameters;
\item the third dredge-up and its efficiency;
\item an assumed composition of the third dredge-up material.
\end{enumerate}
The calculations used here are performed using a mixing length parameter 
$\alpha\,=\,2$, and the mass loss prescription of 
Bl\"ocker \& Sch\"onberner (1993) with a mass loss efficiency of 
$\eta\,=\,0.1$ (Groenewegen \& de Jong 1994).
The end of TP-AGB phase is determined either by the total ejection of 
the stellar envelope, or by the core mass growing up to the Chandrasekhar 
mass limit.

The properties of the TP-AGB stars are sensitive to the initial
metal content. This in particular affects the formation of
carbon stars. The LM2001 models are able to reproduce the
observed relation between the metallicity and the relative
number of carbon stars (Pritchet et al. 1987), in the sense
that lower initial metallicities result in higher proportions
of carbon stars among TP-AGB stars, carbon stars being the dominant 
spectral type for [Fe/H] $\la$ -1 (Blanco et al. 1978, Cook et al. 1986). 
This behaviour reflects the fact that the fraction of the 
total TP-AGB lifetime a model star spends as a carbon rich object is
a function of metallicity. Stars at Z=0.05 with initial masses of
2-3\,M$_{\odot}$ (i.e. the privileged mass range for the formation of 
carbon stars; Mouhcine \& Lan\c{c}on \cite{ML01_Cstars})
spend $\la$\,10\,\% of their total TP-AGB lifetime as carbon
stars, while the fraction reaches $\sim 80\,\%$ for
the metallicity of the LMC or lower.

\begin{figure*}
\includegraphics[clip=,width=\textwidth]{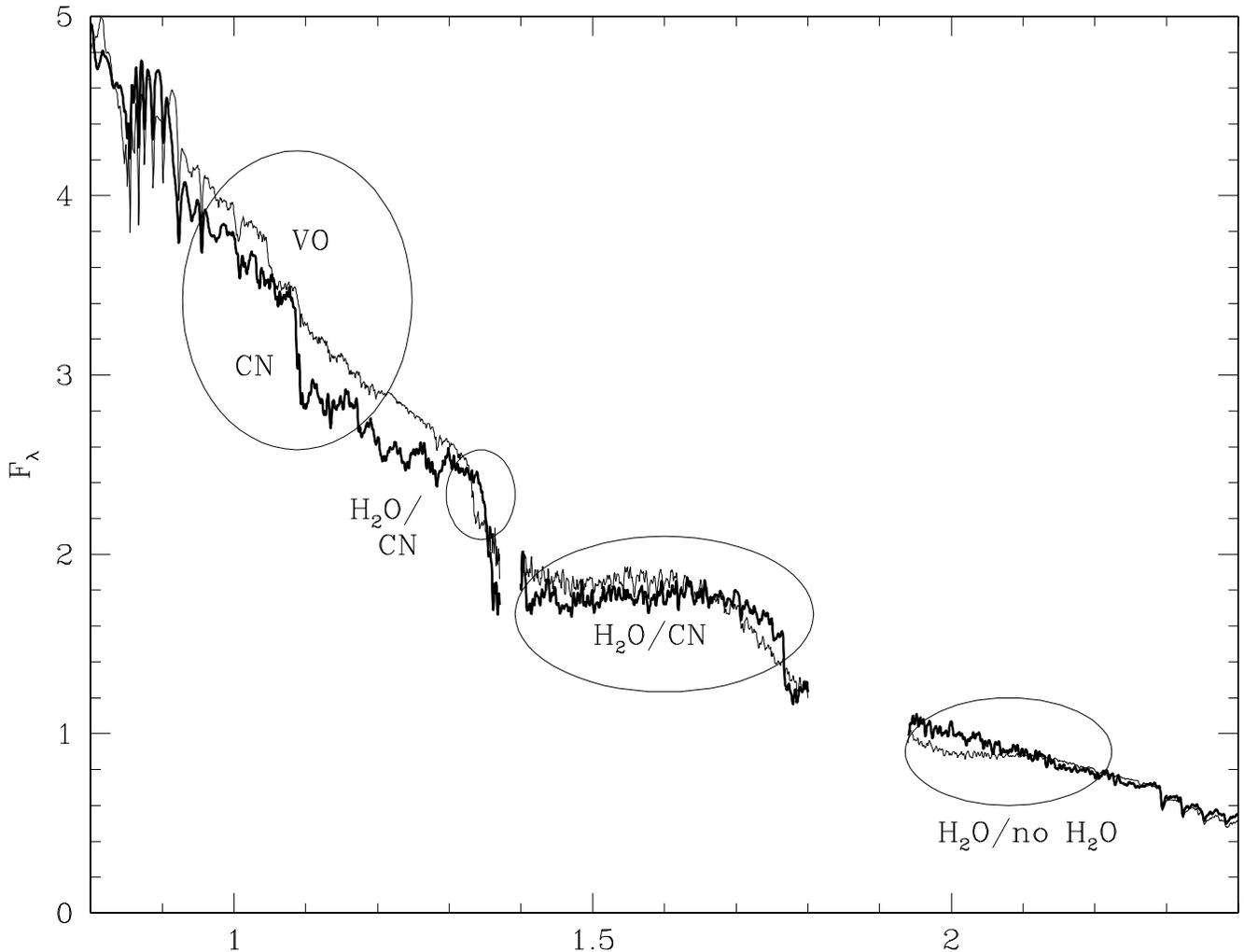}
\caption[]{The effect of carbon stars on the predicted spectrum of 
an intermediate age cluster (Z=0.008, age\,=\,500\,Myr). The flux 
density is given in arbitrary F$_\lambda$ units, but both spectra
correspond to the same total stellar mass. Thick line: standard model; 
thin line: result obtained with O-rich stellar spectra only.}
\label{Crich_Orich.fig}
\end{figure*}

LW2001 use a code derived from {\sc P\'egase} 
(Fioc \& Rocca-Volmerange 1997) for the synthesis of integrated spectra. 
The spectra of Lan\c{c}on \& Wood (2000), averaged as described
by Lan\c{c}on \& Mouhcine (2001), are used for luminous stars with 
spectral types later than K5, and the colour-corrected spectral library 
of Lejeune et al. (1998) is used elsewhere along the stellar tracks.
Cool and luminous TP-AGB stars are also variable stars. This variability 
strongly affects the near-IR energy distribution of those stars. 
They show deep water absorption bands when they are oxygen rich, and 
strong C$_{2}$ and CN absorption bands when they are carbon rich. Hence
the near-IR spectra of intermediate age stellar populations will change 
as the predominant stars change as function of age
and metallicity. The adopted relation between the effective temperature 
of TP-AGB stars and their spectrum is a second parameter that affects 
the shape of the near-IR SSP spectra.

The evolutionary sequences computed for this paper
assume that stars formed in an instantaneous starburst
(star-formation timescale of 1\,Myr or less), with
a Salpeter (1955) power-law initial mass function extending
from 0.1 to 120\,M$_{\odot}$.
\bigskip

To illustrate the effect of
carbon stars on the integrated cluster light, we have
computed additional model sequences in which exclusively
the spectra of O-rich stars were used, 
regardless of the actual chemical composition given
by the evolutionary tracks. The differences
between the two sequences are most clearly visualized
at subsolar metallicity (because carbon stars are
more important) and at ages between
0.5 and 1\,Gyr (because the near-IR contribution of
the TP-AGB stars are largest at these times, in
the ML2002 models). An example is provided
in Fig.\,\ref{Crich_Orich.fig}.

The figure shows that the only {\it{broad band flux}} 
measurement sensitive to the use of C-rich spectra is 
the J band, due to CN and C$_2$ opacities (see Loidl et al. 2001
for band identifications in carbon star spectra, and
Lan\c{c}on \& Wood 2000 for the O-rich features).
The effect on the integrated H and K fluxes is
negligible. Using low resolution spectroscopy implicitely 
accounts for the JHK energy distribution, as long as this 
whole spectral range is covered continuously. The main 
discriminating {\it{spectral features}} can be measured 
at low spectral resolution ($\delta \lambda / \lambda$
of a few 100) or with high signal-to-noise narrow band
photometry (Lan\c{c}on et al. 1999).
When carbon star spectra are included in the models, 
the VO band at 1.05\,$\mu$m vanishes in favour of the CN
bandhead at 1.09\,$\mu$m; the H$_2$O bandhead at 1.33\,$\mu$m
is replaced with a CN bandhead at 1.35\,$\mu$m; the absence
of the wings of the H$_2$O bands on both sides of the H
window give the H spectrum a flat appearance, with an
abrupt cut-off at 1.77\,$\mu$m due to C$_2$; and
for the same reason the K band spectrum is also straight.
\bigskip

In the rest of this paper, our aim is to answer the following questions:  

$\bullet$ Can models for the near-IR spectrophotometric evolution 
of star clusters reproduce the spectrum of W3, with an age, metallicity, 
extinction and mass consistent with constraints from optical spectra\,?

$\bullet$ What constraints are obtained on the chemical nature of the 
AGB stars present in W3\,? 

$\bullet$ Conversely, what constraints does the AGB population of W3 
set on the models for the formation of carbon stars\,? 

$\bullet$ What temperature scales for AGB star spectra are consistent 
with the data\,? 

$\bullet$ How much of this information could have been determined from 
the near-IR SOFI spectrum only, i.e. in the absence of an priori knowledge 
of the age from an optical spectrum\,?.

\section{The Near-IR Spectrum of Cluster W3: the Detection of
Carbon Stars}
\label{Cstar.evidence.sec}

It is well known from counts in the Magellanic Cloud clusters
that carbon stars represent a large fraction of the AGB 
stars of intermediate age clusters, at least at sub-solar 
metallicity (e.g. Aaronson \& Mould 1980, Persson et al. 1983, 
Frogel et al. 1990, Westerlund et al. 1993, Ferraro et al.
1995). Population synthesis models predict the predominance of 
carbon stars over M type stars in the near-IR, under favourable 
age and metallicity circumstances. In agreement with star counts, 
they predict that carbon stars may provide up to 50\,\% of the 
near-IR light (LM2001). 
Nevertheless, carbon stars have never been searched for in galaxies 
with unresolved stellar content. By including {\em spectra} of 
carbon stars in evolutionary spectral synthesis models, 
Lan\c{c}on et al. (1999) opened a new window on distant carbon star 
populations. In this Section, we use NGC\,7252--W3 to validate that 
approach. 

\begin{figure*}
\includegraphics[clip=,width=0.5\textwidth]{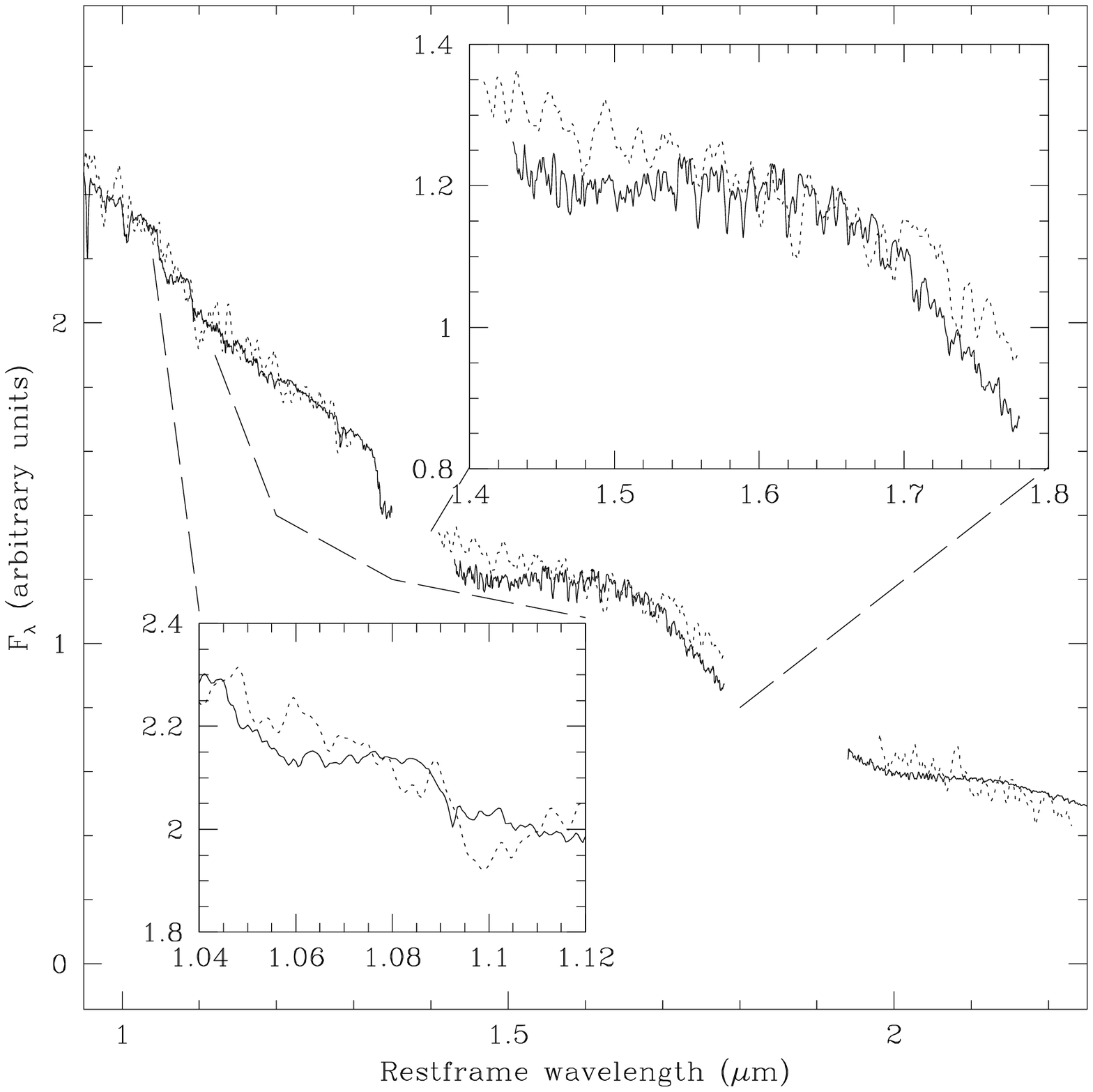}
\includegraphics[clip=,width=0.5\textwidth]{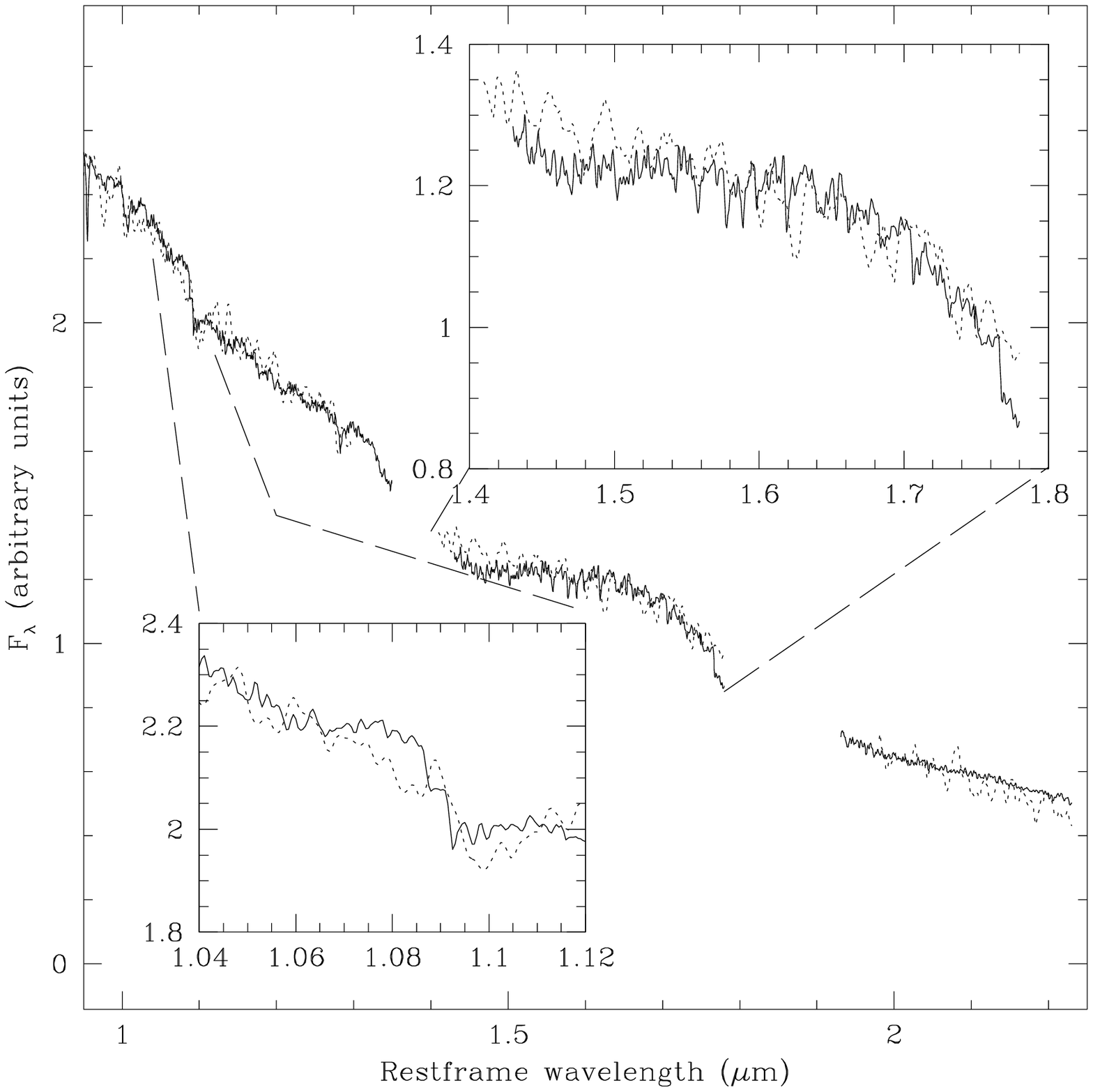}
\caption[]{Evidence for carbon stars in the spectrum of W3. Right 
panel shows comparison between the observed W3 spectrum (dotted-line)
and the synthesis model (continuous line )where carbon stars were 
allowed to form, while the left panel shows a comparison to a 
synthesis model where all TP-AGB stars were forced to remain oxygen 
rich during the whole phase (see Section \ref{Cstar.evidence.sec}). 
The data has been smoothed to a resolution of $\sim 50\,\AA$.}
\label{Cstar.evidence.fig}
\end{figure*}

Are carbon rich stars present in cluster W3, as expected
from the intermediate age cluster models? 
In Fig.\,\ref{Cstar.evidence.fig}, the SOFI spectrum
of W3 is compared with models for a 300\,Myr old
stellar population at solar metallicity. In the right 
frame, carbon star spectra are used when the surface
abundance of carbon exceeds that of oxygen; in the 
left frame, oxygen rich spectra are used exclusively.
The choices of age, metallicity and extinction are
justified in Section \ref{Age_Ext.sec} and \ref{Z_Tscale.sec}.

At 300\,Myr, the contribution of the TP-AGB stars 
to the near-IR light is smaller than at the 500\,Myr
of Fig.\,\ref{Crich_Orich.fig}.
Thus the effect of carbon stars is also less pronounced.
In addition, some key molecular bandhead could not
be recovered because of the strong and variable 
telluric absorption at La Silla.
Nevertheless, in all the remaining spectral regions of 
interest (identified in Fig.\,\ref{Crich_Orich.fig})
the agreement is systematically better when carbon star 
spectra are included. The strongest evidence for carbon
stars comes from the absence of the VO band at 1.05\,$\mu$m,
the presence of the CN bandhead at 1.08\,$\mu$m, and 
the small curvature of the energy distribution inside the
H window. For each of these, the deviation between the smoothed
data and the exclusively O-rich models are at the 3\,$\sigma$
level. Taken together, the signatures provide strong 
evidence for the presence of carbon stars.
The shape of the K window energy distribution also
favours this conclusion, but here the difference between
models is not statistically significant in view of the 
observational noise.

\medskip

The detection of carbon stars in W3 demonstrates that TP-AGB 
studies can be extended to distant, unresolved stellar populations. 
Clearly, better signal-to-noise ratios should be seeked. 
Broadband colours are not sufficient for studies of the 
nature of the TP-AGB stars, in particular when the amount 
of reddening is uncertain. 
The shape of the low resolution near-IR spectrum between
1 and 2.5\,$\mu$m carries most of the relevant information.
It is essential to recover this global shape. 
Spectrographs with a broad wavelength coverage 
and a wide overlap between individual
spectral segments are the most appropriate instruments.
Multiple-filter narrow-band photometry, for instance
with the future refurbished camera NICMOS on board HST, is 
an alternative  approach if the absolute flux levels can be 
calibrated with a high accuracy (Lan\c{c}on et al. 1999). 

The effects of carbon stars on the spectrum of W3 also make 
it clear that near-IR evolutionary synthesis models must
incorporate the spectral contribution of these objects if
they are to be used in the analysis of future high signal-to-noise
near-IR data.  

\section{Age and extinction from the near-IR spectrum and the 
UBVIK$_s$ colours}
\label{Age_Ext.sec}

Figure\,\ref{Cstar.evidence.fig} not only highlights the role of 
carbon stars in W3. It shows that the new spectral synthesis models 
of Mouhcine \& Lan\c{c}on (2002a) are indeed able to reproduce the
integrated spectrum of the cluster, for an intermediate
age and at a metallicity consistent with the estimates
based on optical spectra (Schweizer \& Seitzer 1998, 
Maraston et al. 2001).

In this and the following sSctions, we will examine what 
information on age, extinction and metallicity can be recovered 
in the absence of optical spectroscopy, using only the near-IR 
spectrum, only the UBVIK$_s$ photometry, or these two sets of data 
combined. 

Here, we first focus on age and extinction. As the reddening of  
NGC\,7252 and, more specifically, W3 is poorly known, we consider 
visual extinction as a parameter that needs to be better constrained 
with the new data.

The analysis of the optical spectra of the brightest 
star clusters of NGC 7252 by Schweizer \& Seitzer (1998) 
and Maraston et al. (2001)  has yielded metallicity estimates. 
Schweizer \& Seitzer (1998) favoured solar metallicity. Maraston et al. 
prefer the metallicity of the LMC for W3, but solar abundances 
for other luminous clusters with similar ages. 
Although it is not strictly impossible that coeval clusters 
formed from material with different abundances in a 
merger, we estimate that the apparent spread in
metallicity is more likely to represent uncertainties.
Therefore, this paper will consider both solar and LMC 
metallicities. Here, we mainly work at Z$_{\odot}$, and we 
come back to this choice in Section \ref{Z_Tscale.sec}.

\subsection{The broadband UBVIK$_s$ colours}
\label{chi2_UBVIK.sec}

\begin{figure}
\includegraphics[clip=,width=0.44\textwidth]{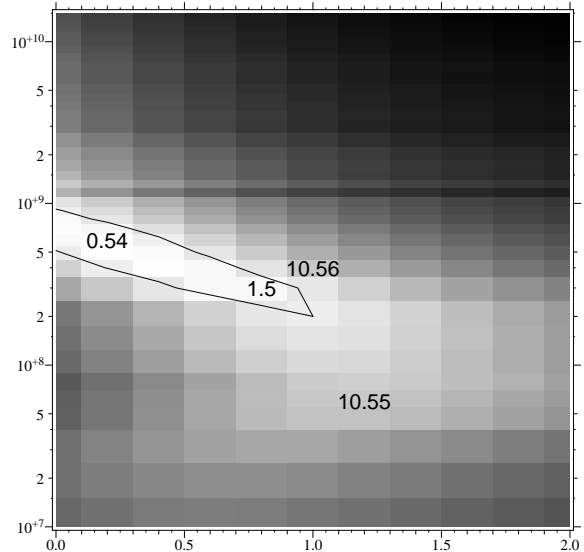}
\caption[]{Goodness-of-fit map obtained in the age-extinction
plane with only the broadband (U-B), (B-V), (V-I) and (I-K$_s$) colours.
Plotted are A$_{V}$ (magnitude) and age (yr) on the X and Y-axis respectively.
The models assume Z=0.02.
The extinction law of Cardelli et al. (1989) with R=3.1 is used. The 
contour delineates the region of good fits (see text).
The minimum $\Delta^2$ value and a few other illustrative values
are indicated.
}
\label{chi2bb_MM_bz02F_R31.fig}
\end{figure}

To assess the relative quality of various model adjustments to the 
data, we have computed goodness-of-fit maps. Each map displays the 
variations of the reduced, weighted quadratic difference between model 
and data (noted $\Delta^2$ hereafter),
for model ages ranging from $10^7$ to $1.5\times\,10^{10}$\,yr and
for model extinctions ranging from A$_V=0$ to A$_V=2$. 
Figure\,\ref{chi2bb_MM_bz02F_R31.fig} shows the typical aspect of
such maps, based on the broad band colours (U-B), (B-V), (V-I) 
and (I-K$_s$) (Miller et al. 1997, Maraston et al. 2001).

The figure displays a rather regular pattern. 
A strong degeneracy between age and extinction is 
apparent: as time passes, the models become 
intrinsically redder, thus requiring less extinction.
The shape of blue young cluster SEDs cannot be
made to match the observed UBVIK$_s$ distribution with any 
amount of obscuration. At solar metallicity, old 
model clusters are intrinsically too red and must also be excluded. 
In agreement with previous studies, only intermediate
ages are found to be acceptable. 

A quantitative interpretation of the maps requires a closer look at the way 
$\Delta^2$ values are computed, and at the sources of uncertainties.
We have: 
$$ \Delta^2 = \frac{1}{4}\ \sum_{i=1}^4\ \frac{(OC_i-MC_i)^2}{\sigma_i^2}$$
where $OC_i$ is an observed colour, $MC_i$ is the corresponding
model colour and $\sigma_i$ is the estimated r.m.s. error on colour $i$.
$\sigma_i$ combines uncertainties in the data and in the models.
For the observational uncertainties, we adopted 
$\sigma=0.06$\,mag for (U-B) and 0.04\,mag for the three other 
colours, as suggested by Miller et al. (1997)
and Maraston et al. (2001).

Miller et al. (1997) have used the transformations of 
Holtzman et al. (1995) to transform HST/WFPC2 colours
into the ``standard" ground-based UBVRI system 
(of Landolt 1983). The UBVI filter passbands of Bessell (1990)
that we use here are closely matched to that standard system, suggesting
passband mismatch errors smaller than 0.02\,mag. When measuring
colours on a model spectrum, we adopt zero colours for a model
of Vega. We have verified that the systematic errors due to this 
convention (Holtzman et al. 1995) and to the  slight differences 
around the Balmer lines between our Vega model and the one used in 
standard HST data reductions (IRAF \verb|stsdas.hst_calib.synphot| package) 
are $\la 0.015$\,mag. 
The K$_s$ measurements of Maraston et al. 
(2001) are tied to the photometric system of Persson et al. (1998), 
who defined the K$_s$ zero point by imposing $K_s=K$ for the 
type A0 standards of Elias et al. (1982).
Systematic errors on (I-K$_s$) are likely to be at least
of the same order as those on optical colours (but are 
more difficult to estimate quantitatively, as empirical
photometric systems focus on either the near-IR
or the optical, not the two combined).
Using 9 tabulations of extinction laws for the Milky Way or
the Magellanic Clouds from the literature, we estimate r.m.s. dispersions
of 0.005 to 0.025\,magnitudes at A$_V$=1, depending on the colour
considered (or more if we include extragalactic extinction laws such as 
given by Calzetti et al 1994, although it is not known if they apply
to individual lines of sight in a galaxy). 
Errors due to stellar evolution prescriptions and 
spectral libraries are difficult to assess precisely.
Based on comparisons between the colours predicted 
by various research groups (e.g. Charlot 1996, Yi et al. 2001, 
Lejeune \& Buser 1999, Salasnich et al. 2000), one must
account for $\sim 0.03$\,mag r.m.s. errors due to the uncertainties
in population synthesis model inputs. The effects of choices in 
the modelling of the TP-AGB are not counted as errors since
they represent the parameters our study must test
(the TP-AGB parameters do not affect (U-B) and (B-V) 
significantly).
In view of all the factors of this paragraph, 
we have accounted for the uncertainties
in the models globally with a factor of 1/2 on $\Delta^2$ (grossly
corresponding to what is expected if the uncertainties on each
model colour equals the uncertainty on the observed colour) and
an extinction dependent correction (corresponding to 0.02\,mag at
A$_V$=1). 

\begin{figure}
\includegraphics[clip=,width=0.44\textwidth]{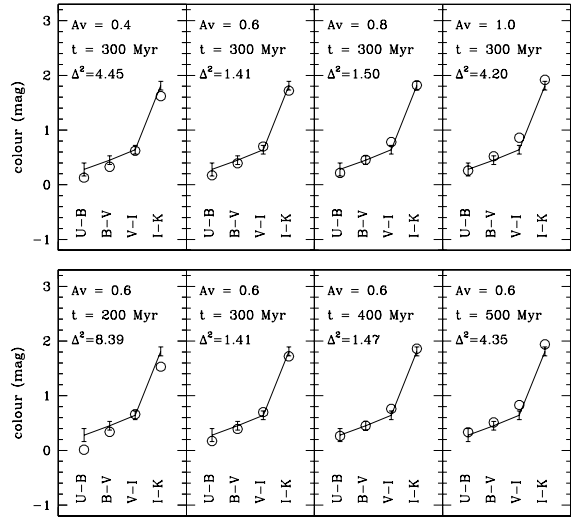}
\caption[]{Comparison between the 4 observed colours (represented as 
$2\,\sigma$ error bars connected with a solid line) and model colours
(open circles).
{\bf Top:} Effect of A$_{V}$ on the broad band colours, at an age of 
300\,Myr. {\bf Bottom:} Effect of age on the broad band colours, for 
A$_{V}=0.6$. The extinction law of Cardelli et al. (1989) 
with R=Av/E(B-V)=3.1 is used. 
The circle sizes are roughly representative of 2\,$\sigma$ errors on 
the models (see text).}
\label{ExtLaw_effect.fig}
\end{figure}

The interpretation of the resulting values in terms of 
agreement probabilities between the data and a model, 
using Pearson's reduced $\chi^2$ probability distribution,
would imply the crude assumptions that the errors have a normal 
distribution and are independent of one another. In particular,
the uncertainties in the model colours are clearly not independent 
(any stellar component contributes to more than one colour, and
extinction affects all colours coherently). Here, we
consider that a model  provides a good fit to the data if
$\Delta^2 < 2$ on the scale of our maps. If our $\chi^2$
estimator indeed followed Pearson's probability distribution
with 4 degrees of freedom, the uncertainties would have a 15\,\% 
probability of producing larger values. Thus our limit is rather 
conservative. Exemples of comparisons between observations and 
models are given in Fig.\,\ref{ExtLaw_effect.fig}.

\subsection{The near-IR spectrum alone}
\label{chi2_NIR.sec}

\begin{figure}[!htb]
\includegraphics[clip=,width=0.44\textwidth]{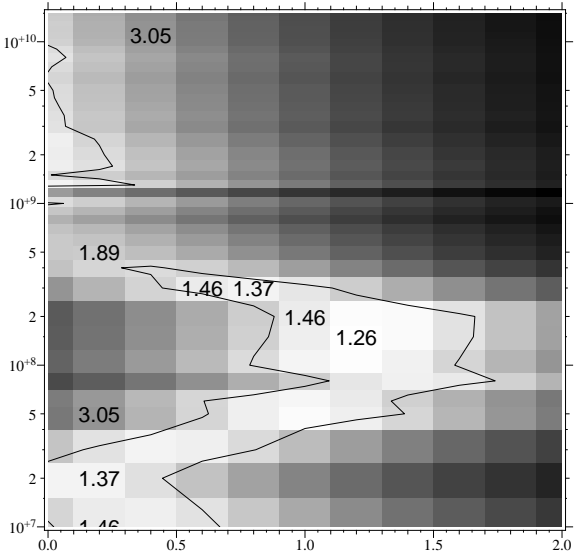}
\caption[]{Goodness-of-fit 
map obtained in the age-extinction plane with the SOFI
spectrum of W3 only (axes as in Fig.\,\ref{chi2bb_MM_bz02F_R31.fig}).
The models assume Z=0.02.
The extinction law of Cardelli, Clayton \& Mathis (1989) with R=3.1
is used. The minimum $\Delta^2$ value and some other illustrative
values are given. The contour indicates regions of acceptable fits.
}
\label{chi2sp_MM_bz02F_R31.fig}
\end{figure}

Figure\,\ref{chi2sp_MM_bz02F_R31.fig} shows the goodness-of-fit map
considering only the near-IR spectrum as constraints.

We have used the r.m.s. deviation $\sigma_{\lambda}$ around a
smoothed continuum as an estimate of the uncertainties in the data.
The squared difference between the model and the
data is weighted with $1/\sigma_{\lambda}^2$ and the sum
of these differences is minimized by applying the adequate
scaling to the model. $\Delta^2$ is obtained by
dividing the resulting sum by
the number of degrees of freedom, $(N-1)$, where $N$ is the
number of pixels used along the spectrum.
We note that we do not correct for the size of the spectral PSF, which
introduces correlations between errors on 2-3 adjacent pixels.
Observational uncertainties on the slope of the spectrum
affect estimates of A$_V$. A 5\,\% uncertainty on the ratio
of the fluxes at 1\,$\mu$m and at 2\,$\mu$m translates into
an uncertainty of 0.15 magnitudes in A$_V$.

Again, the numerical values of $\Delta^2$ cannot be
directly transformed into acceptance probabilities.
The various sources of uncertainties that we have cited
and the eye inspection of a variety of model adjustments
leads us to set the limit of acceptable $\Delta^2$ values
at 1.5. The contour in Fig.\,\ref{chi2sp_MM_bz02F_R31.fig}
identifies the corresponding region of acceptance.
The model adjustment shown in Fig.\,\ref{Cstar.evidence.fig}\,(b)
corresponds to [t=300; A$_V$=0.7]. It provides a good fit to the
data with a $\Delta^2$ of 1.39 on Fig.\,\ref{chi2sp_MM_bz02F_R31.fig}.
\medskip

A curved line of relatively low $\Delta^2$ values runs through 
the map, reflecting the degeneracy between extinction and 
stellar evolution in producing a red enough broad band energy 
distributions between 1 and 2.5\,$\mu$m. Three main segments 
can be identified along this curve. (i) At ages between $10^7$ 
and $10^8$\,yr, red supergiants are the main sources of near-IR 
light. At about 20\,Myr, they provide  intrinsic {\em near-IR} 
colours that are similar to those of W3 even without much extinction.
Later on, stars on extended blue loops of the evolutionary
tracks make the integrated colours bluer, shifting the 
$\Delta^2$-valley towards higher extinctions.
(ii) In the intermediate age
regime, the intrinsic near-IR colours become redder with time,
as the relative contribution of cool AGB stars increases.
The need for extinction is progressively reduced. The poor 
fit between 500\,Myr and $\sim 1$\,Gyr is due to the strength of
molecular features in the model spectra at these ages. In the
observed spectrum, these features are present, but with
a relatively small contrast.  (iii) At 
old ages ($> 10^9$\,yr), the near-IR model spectra are intrinsically
as red as or redder than the data and any extinction would
worsen the agreement.

The minimum $\Delta^2$ values are obtained at relatively
young intermediate ages (150-200\,Myr) with relatively
high values of the extinction (A$_V=1-1.4$, i.e. values 
consistent with those obtained for the H\,II region
S101 of NGC\,7252 by Schweizer \& Seitzer 1998).
The exact location of the minimum depends on the wavelength
dependence of the weighting adopted in the $\Delta^2$ 
computation and on the extinction law. For instance, varying the 
ratio R=A$_V$/E($B$-$V$) from 2.7 to 5 (as observed along
various lines of sight in the Milky Way; Cardelli et al. 1989)
shifts the minimum $\Delta^2$ from [t=200; A$_V$=1.4] to
[t=150; A$_V$=1.0], without modifying the actual value of 
the minimum $\Delta^2$ significantly.

\subsection{The broadband colours and the near-IR spectrum combined}

The comparison of Figs.\,\ref{chi2sp_MM_bz02F_R31.fig} and
\ref{chi2bb_MM_bz02F_R31.fig} shows that, in the case of cluster
W3, the degeneracy between age and extinction can be 
broken with the simultaneous use of the UBVIK$_s$ colours
and the 1--2.3\,$\mu$m spectrum. Synthetic near-IR spectra
at ages of 500--700 Myr, with low extinction optical depths,
provide satisfactory UBVIK$_s$ colours but the molecular bands
they display are too pronounced to be acceptable.
As a matter of fact, the observed molecular bands are rather 
shallow. Models younger than 250\,Myr are rejected because of
their UBVIK$_s$ colours. The overlap of the contoured regions
of the two figures identifies the models that comply with
all the constraints.
Our prefered solar metallicity model corresonds to
[t=300\,Myr; A$_V=0.8$]. Taking into account the various sources
of errors mentioned above (extinction laws, weighting of the
spectral elements, photometric filter passbands and zero points),
acceptable models lie between [t=200\,Myr; A$_V=1.0$] and
[t=400\,Myr; A$_V=0.5$].

\begin{figure}
\includegraphics[clip=,width=0.44\textwidth]{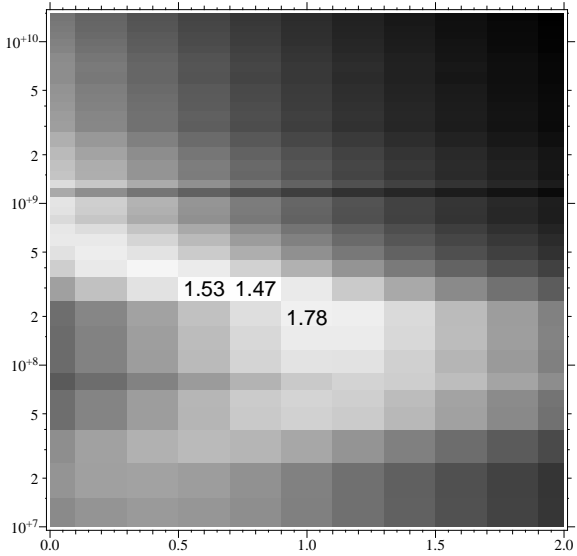}
\caption[]{$\Delta^2$ map obtained in the age-extinction plane considering 
simultaneously the broad-band photometry and the near-IR spectrum 
of W3 cluster (Z=0.02, extinction law as in Fig.\,\ref{chi2bb_MM_bz02F_R31.fig}
and \ref{chi2sp_MM_bz02F_R31.fig}).}
\label{chi2all_MM_bz02F_R31.fig}
\end{figure}

Fig.\,\ref{chi2all_MM_bz02F_R31.fig} shows a goodness-of-fit map 
constructed as a linear combination of the ones previously discussed. 
The weights are chosen such as to give the set of 4 broad-band 
colours and the large set of near-IR flux densities equal importance
in the combined $\Delta^2$.
Maps of this type will allow us to compare model adjustments when 
varying internal parameters such as the metallicity or the temperature 
of the input LPV spectra.

\section{Constraints on metallicity and on the temperature
scales of LPV spectra}
\label{Z_Tscale.sec}

Previous studies of the clusters of NGC\,7252 suggested
metallicities between Z$_{\odot}$ and Z$_{\rm LMC}$ (see 
Section \ref{selection.sec}). Our models favour a solar
metallicity for cluster W3, although LMC metallicity
cannot be safely rejected.

\begin{figure}
\includegraphics[clip=,width=0.44\textwidth]{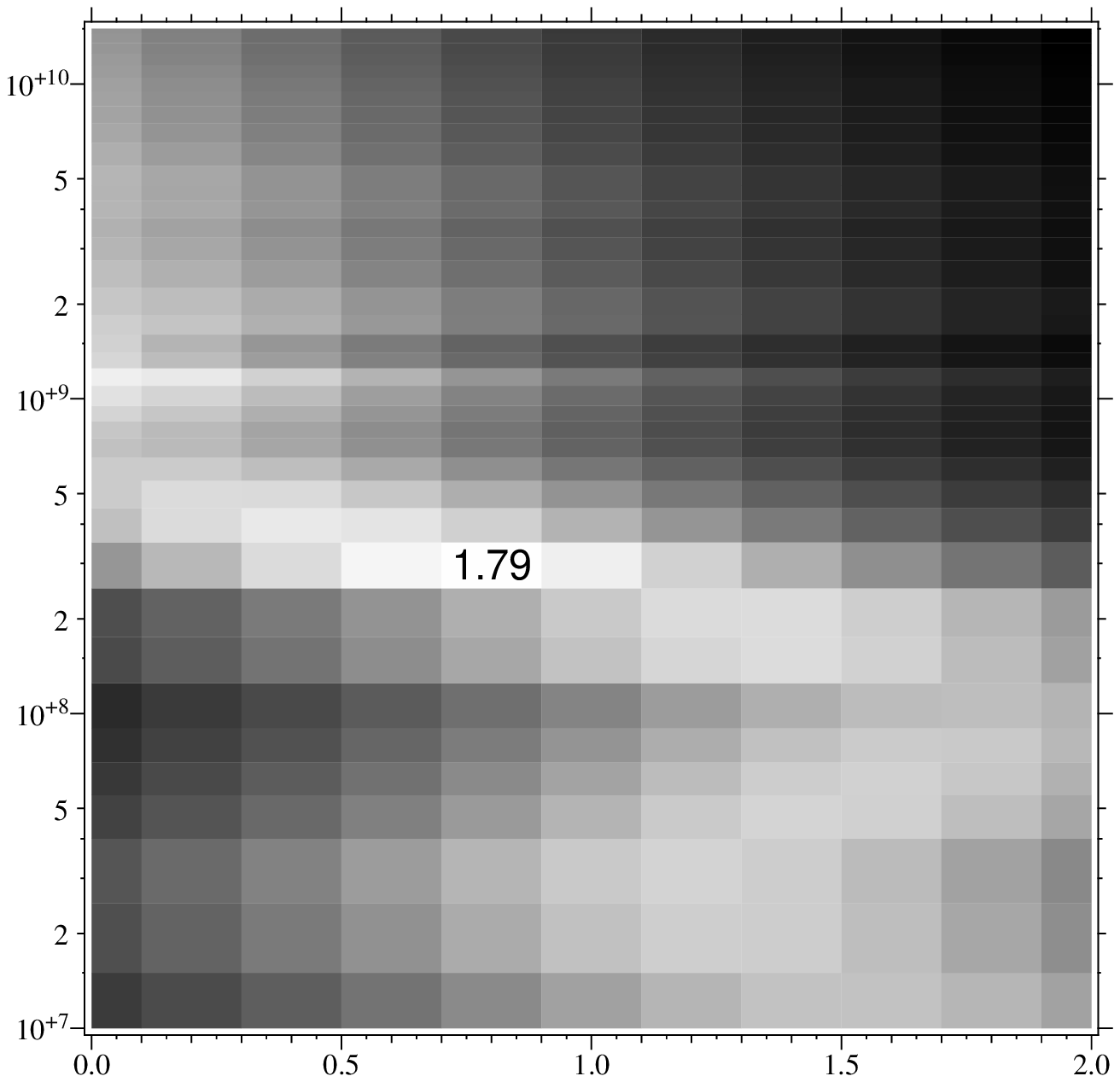}\\
\includegraphics[clip=,width=0.44\textwidth]{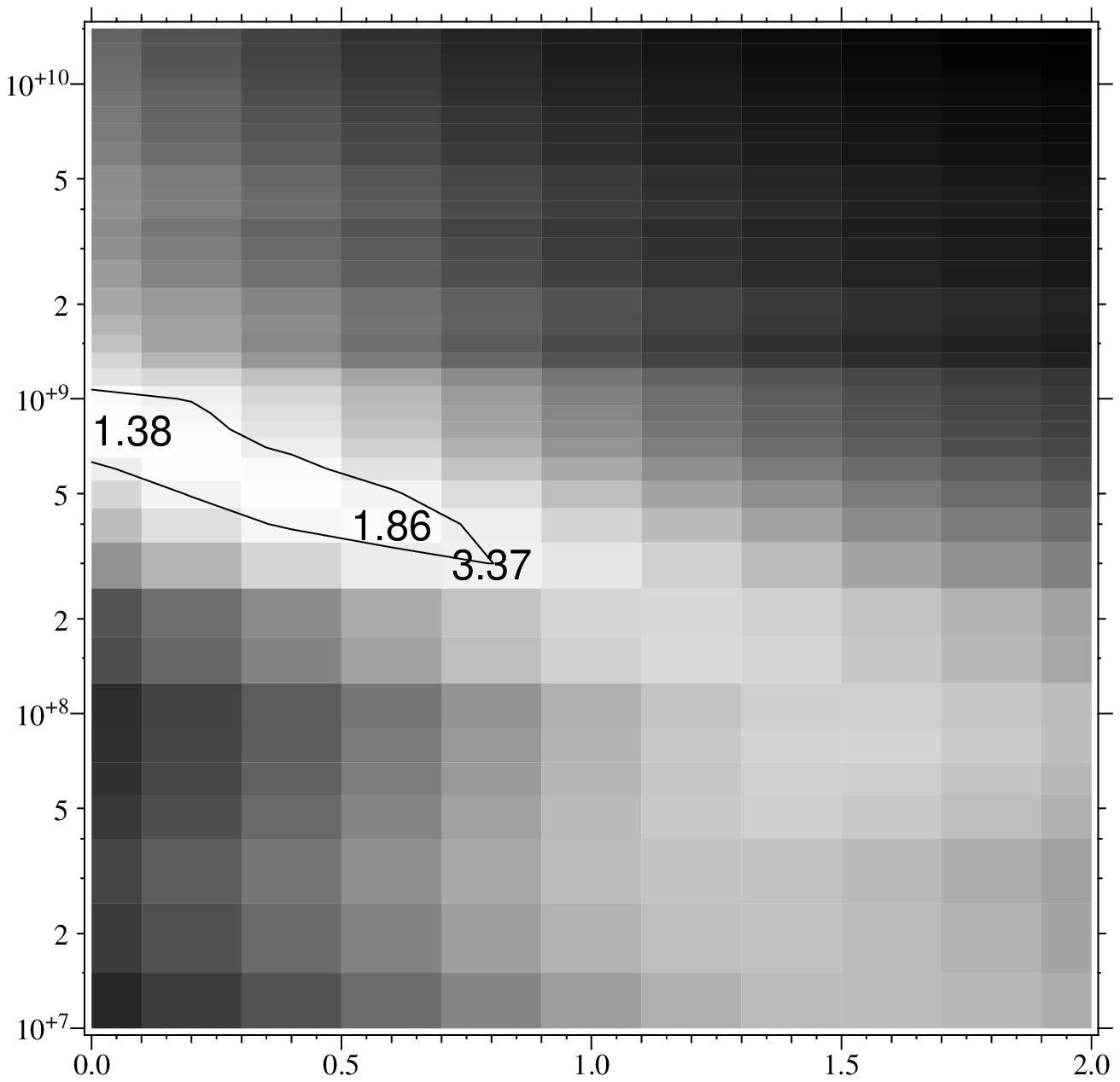}\\
\includegraphics[clip=,width=0.44\textwidth]{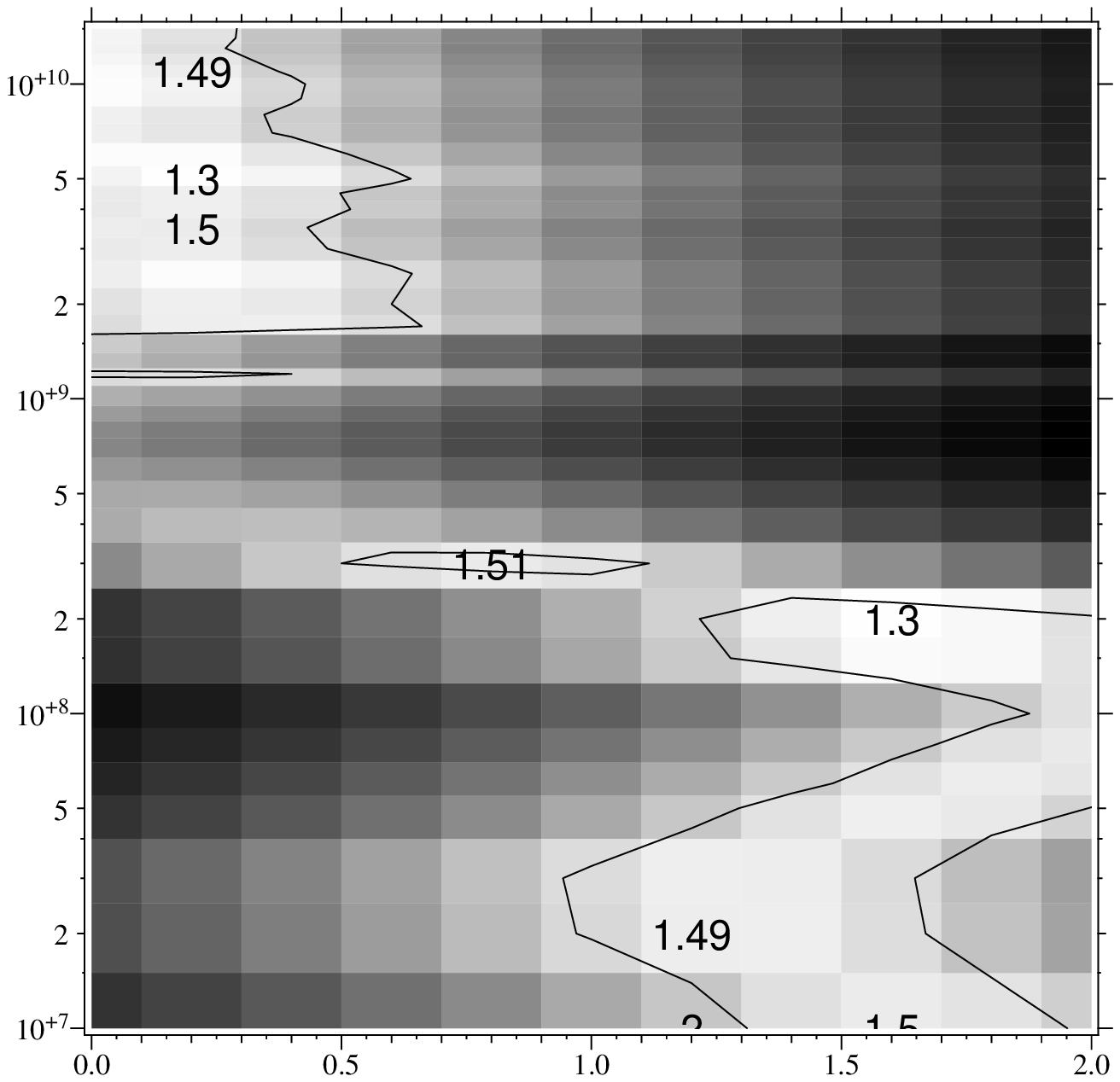}\\
\caption[]{Goodness-of-fit maps for models at Z=0.008. The layout
and the $\Delta^2$ computations are as in 
Figs.\,\ref{chi2all_MM_bz02F_R31.fig}, \ref{chi2bb_MM_bz02F_R31.fig}
and \ref{chi2sp_MM_bz02F_R31.fig}. {\bf Top:} broad band colours and
SOFI spectrum combined; {\bf middle: } colours only;
{\bf bottom:} SOFI spectrum only.
}
\label{chi2_z008.fig}
\end{figure}

The combined goodness-of-fit map, an LMC-metallicity
equivalent of Fig.\,\ref{chi2all_MM_bz02F_R31.fig},
has a very similar appearance, but the minimum value 
is 1.79 instead of 1.47 at Z$_{\odot}$ 
(cf Fig.\,\ref{chi2_z008.fig}, top). The location of 
the minimum is unchanged: [t=300\,Myr; A$_V$=0.8].

The comparison of the middle plot of Fig.\,\ref{chi2_z008.fig} with 
Fig.\,\ref{chi2bb_MM_bz02F_R31.fig} suggests that {\it{at a given age}},
the intrinsic colours of a stellar population are globally bluer
when the stellar metallicities (and therfore opacities) are reduced 
and a higher value of A$_V$ compensates for this effect. 

At [t=300\,Myr; A$_V$=0.8], both the near-IR and the UBVIK$_s$ 
$\Delta^2$ values are slightly higher than those obtained at
solar metallicity. In the near-IR, the difference is not
significant. The difference in the UBVIK$_s$ map is more 
important and according to our discussion in 
Sec.\,\ref{chi2_UBVIK.sec} $\Delta>2$ is not an acceptable
fit. However, a reasonable change in the adopted extinction 
law produces acceptable UBVIK$_s$ colours. For instance, 
$\Delta^2$ is reduced from 3.37 to 1.6 when we use the 
extinction law of Cardelli et al. (1989) with R=A$_v$/E(B-V)=5. 
All creteria for a good fit are met at [t=300\,Myr; A$_V$=0.8. 
\bigskip

At solar metallicity, a significant fraction of the TP-AGB stars 
are oxygen rich. Assuming a solar metallicity, we may thus attempt
to constrain the temperature scale of the M type LPV spectra:
the predictions of near-IR spectrophotometric 
properties of intermediate-age stellar populations (mainly 
narrow-band molecular indices) show a rather strong dependence upon this 
calibration (Lan\c{c}on et al. 1999, Mouhcine \& Lan\c{c}on 2002). 
If one affects high effective temperatures to stellar spectra 
with deep molecular features in the input stellar library of the
models, one produces integrated spectra that also have 
pronounced molecular features.
As discussed by Lan\c{c}on \& Mouhcine (2002), conclusions on such a 
calibration will be model dependent, in the sense that they are strongly
linked to the way the effective temperatures of the evolutionary tracks
have been computed.

\begin{figure}
\includegraphics[clip=,width=0.44\textwidth]{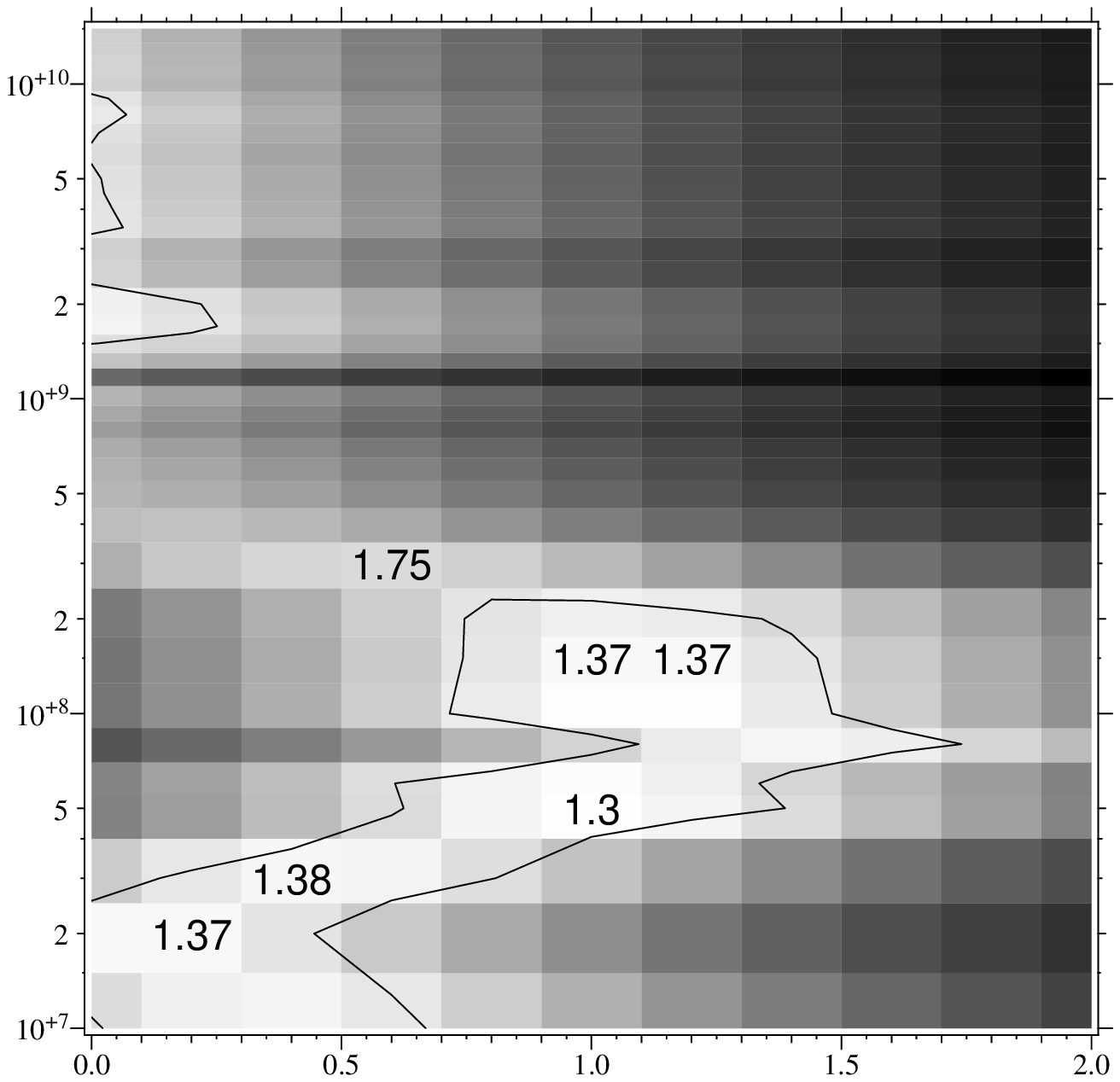}\\
\includegraphics[clip=,width=0.44\textwidth]{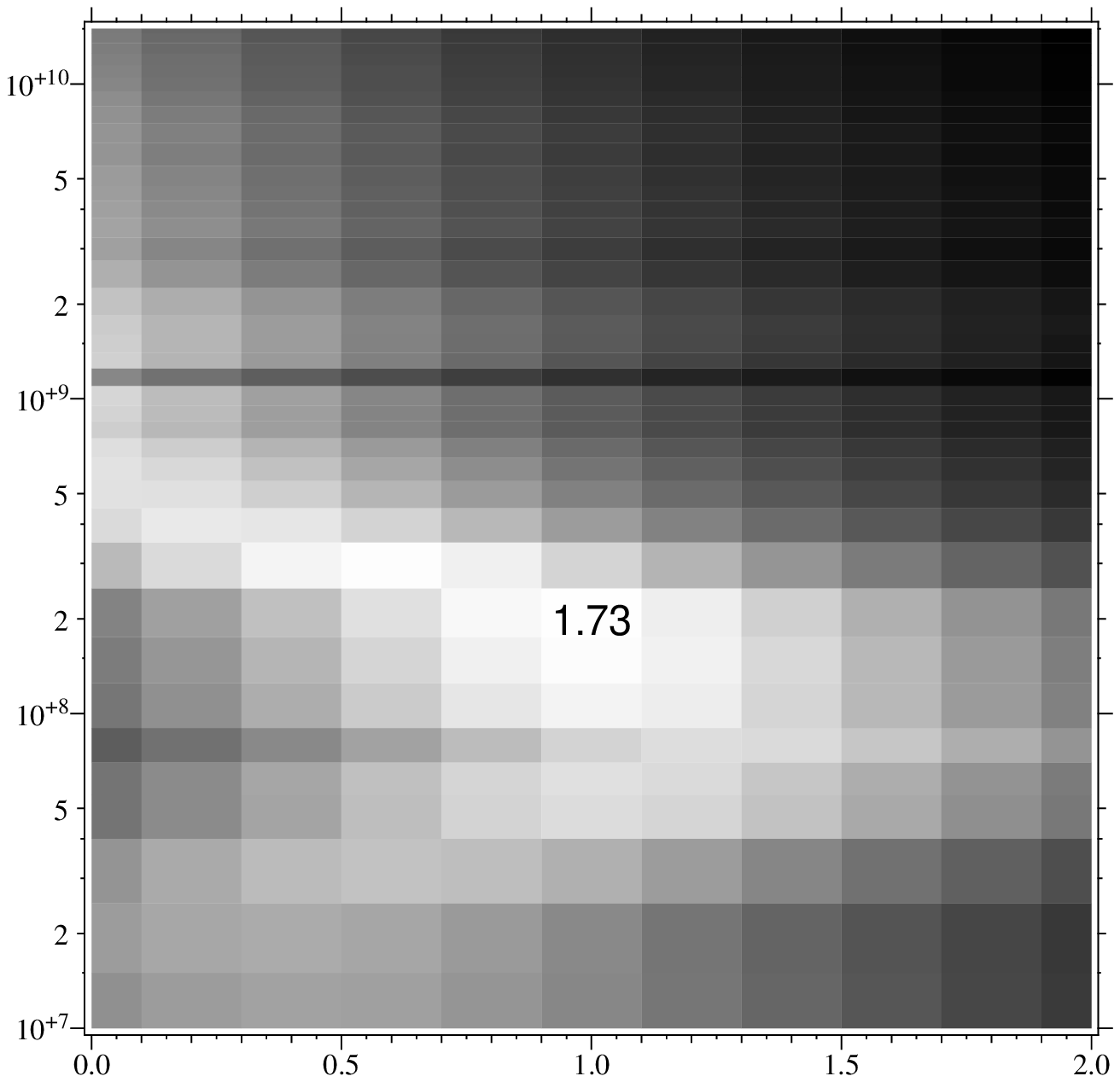}
\caption[]{Goodness-of-fit map obtained in the age-extinction plane for solar 
metallicity models using the Bessell et al. (1989a) effective temperature 
scale for TP-AGB oxygen rich stars. Top: only the near-IR spectrum 
of the W3 cluster is considered. Bottom: both the broad-band photometry and 
the near-IR spectrum are considered.}
\label{chi2.v9_MM.bz02.all.fig}
\end{figure}

Two different effective temperature scales for LPV spectra may be 
considered. The first one is that LPV stars may have the same 
effective temperature scale as static giant stars; the 
second one is that the effective temperature scale of LPVs may 
differ from that of the static giants. 
In principle one may expect that, for the same effective temperature, 
a variable giant have redder colours than its parent static 
giant star, because pulsation makes the atmosphere of the star 
more extended, and consequently redder (Bessell et al. 1989b).  

We have constructed two grids of models assuming two different 
effective temperature scales for the LPV stars. The first one is 
taken from Bessell et al. (1989a) who consider only static
M stars (though already with extended atmospheres). 
The authors suggest that the proposed scale is appropriate for AGB 
stars. The second effective temperature scale is one proposed by 
Feast (1996), based on angular diameter measurements for late-type 
stars combining Miras and non-Mira M-type stars. 
The second temperature scale is much steeper than the first one
at the low temperatures relevent for TP-AGB stars; smaller
decreases in temperature are needed to explain the same spectral
evolution to redder colour and deeper molecular bands.
For the evolution of integrated properties of stellar populations, 
this means that models calculated with the temperature scale
of Bessell et al. (1989a) will produce integrated stellar population 
spectra with more pronounced near-IR molecular features than 
those calculated assuming the scale of Feast (1996).

The goodness-of-fit map for models that assume the static giant 
star scale from Bessell et al. (1989a), combining the 
constraints from the broad-band photometry and the near-IR spectrum, 
is shown in Fig.\,\ref{chi2.v9_MM.bz02.all.fig}. All previous 
figures assumed the effective temperature scale of Feast (1996).
The adopted temperature scale has almost no effect 
on the evolution of broad-band photometry; the goodness-of-fit 
map obtained comparing broad-band colours to models constructed 
assuming Bessell et al. (1989a) scale is similar to 
Fig.\,\ref{chi2bb_MM_bz02F_R31.fig}, obtained using the scale of 
Feast.
The map shows that adopting the temperature scale of Bessell et al. 
for the O-rich spectra degrades the fits. 
This temprature scale assigns a higher effective temperature 
to a given spectrum than the scale of Feast. 
Thus, the use of spectra with strong molecular bands is extended to 
higher temperature regions in the HR diagram, and molecular features
are also stronger in the integrated spectrum of the population. 
As a result, the VO band (1.05\,$\mu$m) is present at ages as young 
as 300\,Myr and the wings of the H$_2$O bands curve the H and K-band
energy distributions (in a way similar to what is shown in 
Fig.\,\ref{Cstar.evidence.fig}.

\section{The influence of the TP-AGB evolutionary tracks}
\label{comp.sec}

As discussed in Sec.\,\ref{models.sec}, the treatment of the 
TP-AGB evolutionary phase in population synthesis models is
a major challenge. The predicted evolution of intermediate-age 
stellar populations (mainly for ages in the range of 0.1 to 
$\sim\,1.5$ Gyr) from different sets of models show large 
differences. In this Section we will compare our results 
with those obtained with other models that use a different
modelling of the TP-AGB evolutionary phase. 

We have compared the observed spectra of W3 to the model sets of 
Lan\c{c}on et al. (1999). These models were calculated using the 
prescriptions of Groenewegen \& de Jong (1993) for the TP-AGB 
phase. Note that the effect of the envelope burning on the 
luminosity of TP-AGB stars was negelected in these models. 
The predicted evolution of broad-band colours by 
Lan\c{c}on et al. (1999) models is similar to the prediction 
made by the recent version of Bruzual \& Charlot models as cited 
in Maraston et al. (2001), at least regarding (V-K) and (B-V). 
For the stellar populations in the age range of $0.1-1\,$Gyr,
the age interval into which falls W3, the  models of
Lan\c{c}on et al. (1999) predict that the reddest 
optical/near-IR or near-IR/near-IR colours are reached 
at ages of $\sim\,200$ Myr. The colours become bluer afterwards,
up to 1\,Gyr.
This discrepancy relative to our present models, where the 
reddest colours are obtained for stellar populations at ages 
$\sim\,0.8-1\,$Gyr, is due to the fact that the TP-AGB lifetime 
of massive TP-AGB stars, and hence their contribution to integrated 
light, is overestimated when the overluminosity due to envelope
burning is neglected. Note that the new models reproduce
the evolution of the contribution of the AGB stars to the 
bolometric luminosity as a function of age, as observed in LMC 
clusters (Frogel et al. 1990).

We find that Z=0.008 models from Lan\c{c}on et al. (1999) only 
provide marginal matches to broad-band photometry with any extinction 
or age ($\Delta^2 \simeq 4$). At solar metallicity, the UBVIK$_s$ 
can be matched well, but not without extinction. 
The ``best fit" found with this model set is at [t=300--400; A$_V$=0.6] 
again. Indeed, the ML2002 models and those of Lan\c{c}on et al. (1999)
produce very similar spectra at an age of 300 Myr.\\
Maraston et al. (2001) find young photometric ages even with the 
assumption of negligible extinction. Comparisons of their models and 
ours show clear discrepancies. Their models predict redder 
colours than ours at the same age. At ages between 0.2\,Gyr and 
1\,Gyr, (B-V) predicted by Maraston et al. (2001) is redder than ours 
by 0.15 magnitudes or more, their (V-K) is bluer than ours by 0.1 mag. 
at 0.2\,Gyr, and redder by 0.1 mag. at 1 Gyr. At the approximate age 
of W3, a change of 0.15 mag in (B-V) corresponds to a change of 
$50-60\,\%$ in age. 
The relative young photometric ages derived by Maraston et al. (2001)
can be explained partly by the differences in the (B-V) evolution. 
The inclusion of the TP-AGB stars in both models is quite different, 
and may be another source of the differences between their and our 
results for the cluster photometric age. 
At fixed age, their models predict a higher contribution of the AGB 
stars to the K-band light than ours. Indeed, at the approximate age of 
W3, our models predict that AGB stars contribute $\sim\,45\%$ to the 
integrated K-band light, while the models used by Maraston et al. (2001) 
predict that these stars contribute of $\sim\,55\%$. 

As noted in the previous Sections, our best fit has an age of 
300 Myr and A$_V$=0.7-0.8. If we had used only UBVIK$_s$ photometry 
as constraints, we would have favoured older ages (600-800\,Myr) 
and lower extinction values (A$_V$=0--0.3). 
This would have been consistent with the early photometric
age estimates of Miller et al. (1997), who assumed low dust 
optical depths. Optical spectroscopy leads to younger ages 
(300-600 Myr: Schweizer \& Seitzer 1998; 250-300 Myr, 
Maraston et al. 2001). Our ages agree with previous spectroscopic 
ages, which are independent of extinction.\\

\section{Summary and conclusions}
\label{concl}

Several of the star clusters in NGC 7252 are sufficiently massive 
to overcome the problem of the large stochastic fluctuations due 
to the small number of bright AGB stars in intermediate-age star 
clusters in nearby galaxies. This gives the unique opportunity to 
test the accuracy of theoretical modelling of intermediate-age
populations in the near-IR. With this aim, we have obtained near-IR 
spectroscopic and photometric observations of star clusters in 
the merger remnant NGC 7252. 

We used new models for simple stellar populations, in which the 
contribution of AGB stars to the integrated light is treated in 
a consistent way. We were able to match the relevant observational 
constraints on the intermediate-age AGB stellar population.

UBVIK$_{s}$ photometry imposes the global range of ages but with a 
degeneracy between age and extinction. The absence of independent 
extinction value determinations for the target has a significant effect
on model rejection criteria (at A$_V>0.5$). Higher metallicities move 
solutions to older ages and/or higher extinction values.

At intermediate ages, the strength of the molecular bands is a
discriminant. Fitting the near-IR spectrum of the W3 cluster 
shows that carbon stars are needed to obtain an acceptable fit of the 
data. Considering only oxygen rich stars as representative of the 
whole AGB stellar populations leads to synthetic spectra with spectral
features that are not compatible with the observations. 
This is the first clear indication of the presence of carbon stars 
outside Local Group galaxies. Our models show that the available 
UBVIK$_{s}$ photometry is not sufficient to detect 
carbon stars in the star cluster system of NGC 7252. Using the 
strength of the molecular bands as discriminator, we derived constraints 
on the W3 age: strong TP-AGB features appear after $\sim$ 300 Myr, 
weaker ones before. W3 is intermediate. We mention that the low S/N of 
our data is a severe limitation of the diagnostic power of the near-IR 
spectral features. Spectro-photometric data are consistent with a 
metallicity $0.4-1\,$Z$_{\odot}$ for W3, with higher likelihood for 
solar metallicity. 

Contrary to previous studies of star cluster system of NGC 7252 that 
assumed negligible extinction, we found that good fits of the data require 
high extinction independently of metallicity (i.e., A$_{V}=0.8$).
Using the near-IR constraints has allowed us to break the age-extinction
degeneracy, to show that the assumption of low extinctions was incorrect, 
and to find ages consistent with optical spectroscopic (extinction-
independent) ages.

The comparison of the observed near-IR spectrum of cluster W3 to 
grids of models calculated using our evolutionary tracks with various 
TP-AGB effective temperature scales, shows that the data are better 
reproduced when an effective temperature scale calibrated 
on LPV stars is used (i.e. steeper than the effective temperature scale 
of static giant stars). 

\begin{acknowledgements}
Thanks to Dr P. Francis to provide us the dome flats and the bad pixels 
mask used for the data reduction.  
\end{acknowledgements}

\end{document}